\documentstyle[12pt,epsf]{article}
\newcommand{\mysection}{\setcounter{equation}{0}\section}

\def\beq{\begin{equation}}
\def\eeq{\end{equation}}
\def\beqa{\begin{eqnarray}}
\def\eeqa{\end{eqnarray}}

\begin{document}

\begin {flushright}
EDINBURGH 98/9\\
LBNL-41993
\end {flushright}
\vspace{3mm}
\begin{center}
{\Large \bf Soft-gluon resummation for heavy quark production in 
hadronic collisions}
\end{center}
\vspace{2mm}
\begin{center}
Nikolaos Kidonakis\footnote{Present address:
Department of Physics, Florida State University, Tallahassee,\\
FL 32306-4350}\\
\vspace{2mm}
{\it Department of Physics and Astronomy\\
University of Edinburgh\\
Edinburgh EH9 3JZ, Scotland, UK} \\
\vspace{4mm}
Ramona Vogt\\
{\it Nuclear Science Division} \\
{\it Lawrence Berkeley National Laboratory} \\
{\it Berkeley, CA 94720} \\
{\it and}\\
{\it Physics Department}\\
{\it University of California at Davis} \\
{\it Davis, CA 95616}\\
\end{center}

\vspace{2mm}

\begin{abstract}

We discuss the heavy quark production cross section
near partonic threshold in hadronic collisions, including the 
resummation of leading and next-to-leading logarithms arising 
from soft gluon emission.
We show how to handle the complications due to the non-universal
non-leading logarithms.  We give analytical results for the 
$q {\bar q}$ partonic subprocess and numerical results in the DIS scheme
for top quark 
production at the Fermilab Tevatron where the $q {\bar q}$ channel dominates.

\end{abstract}

\vspace{2mm}
\begin{center}
\end{center}

\newpage

\mysection{Introduction}

Heavy quark production has been a topic of intense interest in the
last few years, particularly since the discovery of the top quark at the
Fermilab Tevatron. The top quark production cross section will be measured
with increased precision as Tevatron running continues.
It is therefore important to make precise theoretical predictions
for the cross section.  Calculations of the cross sections 
for processes such as top quark production
are based on the factorization theorems of perturbative quantum chromodynamics
(pQCD) \cite{CSS}.  Factorization separates the perturbative, short-distance
hard scattering from nonperturbative universal parton distribution functions. 
The hadronic cross section is then given by the convolution of the 
perturbatively calculable partonic cross section with 
the experimentally determined parton densities.
Complete calculations of 
heavy quark production have been carried out up to next-to-leading order (NLO)
in the strong coupling constant $\alpha_s$ \cite{heavycalcs}
and the NLO corrections were found to be significant.
Calculations to higher orders are formidable and full results do not
exist beyond NLO. 
However, near the final-state production threshold, $z \equiv Q^2/s=1$,
where $Q$ is the invariant mass of the heavy quark pair and $s$ is the square
of the center-of-mass energy of the partonic collision,
there are large logarithms at each order
in the perturbative expansion originating from soft gluon emission
which can be resummed to all orders in pQCD. 
At $n$th order in perturbative QCD one encounters terms as singular as
$(-\alpha_s^n/n!) [\ln^{2n-1}((1-z)^{-1})/(1-z)]_+$, which, when folded
with parton distributions, give large and positive corrections \cite{Thuile}.
These Sudakov logarithms, arising from the incomplete cancelation between 
real and virtual corrections near threshold
where the energy of the radiated gluon approaches zero, 
increase the NLO cross section.

Resummation is a result of factorization \cite{CLS} and is most 
easily derived in terms of
moments of the cross section with respect to the variable $\tau=Q^2/S$,
where $S$ is the center-of-mass energy of the incoming hadrons. 
In moment space the hadronic cross section becomes the
product of moments of the parton distributions and the partonic cross section.
The physical resummed cross section is obtained upon inversion of the
exponentiated moments back to momentum space.

The leading logarithms for heavy quark production arise from soft
gluon emission from the quarks and gluons in the incoming hadrons and
are therefore universal. They are thus the same as in the Drell-Yan process 
where Sudakov resummation was applied some time ago \cite{oldDY,CT}.
More recently, leading-log resummed calculations for heavy quark production
\cite{LSN,distr,HERAB,bc,cat} have been presented, using different methods 
\cite{LSN,bc,cat}
to invert the moment calculation to momentum space. The differences
between the results are at the level of subleading logarithms and are
numerically small for top quark production at the Tevatron.
 
The theoretical framework for the extension of resummation to non-universal 
next-to-leading logarithms (NLL) for heavy quark hadroproduction has been 
presented in Refs. \cite{Thuile,NKGS,diss} in moment space, and has recently 
been extended to dijet production \cite{KOS}, and single-particle
inclusive cross sections \cite{LOS}, including direct photon 
production \cite{LOS} and heavy quark electroproduction \cite{MoLa}. 
The formalism has also been used to study dijet rapidity gaps \cite{OS}.
Recently a different method for the
calculation of the total NLL resummed heavy quark cross section has appeared
\cite{BonCat} which agrees with the moment space results
in Refs. \cite{Thuile,NKGS,diss}, and has also been applied to direct photon
production \cite{CMN}.

Beyond leading logarithms, 
the color exchange in the hard scattering must be taken into account.
In the next section we will present an anomalous dimension matrix,
explicitly calculated in Refs.~\cite{NKGS,diss}, which controls 
color-sensitive gluon radiation into the final state.
Using the results in \cite{NKGS} we have previously calculated the NLL 
resummed hadronic cross sections for top and bottom quark production at 
fixed center-of-mass scattering angle \cite{NKJSRV}.  In this paper,
we present analytical results for the angle-integrated NLL resummed 
cross section in the $q {\overline q} \rightarrow Q {\overline Q}$ channel
in the DIS scheme.
The calculation of $gg \rightarrow Q {\overline Q}$ is similar
but more involved and will be discussed in detail elsewhere \cite{NKRV}. 
In Section 3, we give numerical results for top quark production
at the Fermilab Tevatron in the 
$q {\overline q} \rightarrow t {\overline t}$ channel, which is dominant.

\mysection{Resummation formalism}

The resummation of the cross section in moment space is
achieved by refactorizing the partonic cross section into hard components
which describe the truly short-distance hard-scattering, 
center-of-mass distributions associated with gluons
collinear to the incoming partons, and a soft function associated with 
non-collinear soft gluons.  
The soft function is a matrix in color space which satisfies 
a renormalization group equation whose solution provides a matrix 
evolution equation, in terms of soft anomalous dimension matrices, 
that controls threshold logarithms ~\cite{NKGS}. The relevant anomalous
dimension matrices were calculated in Refs.~\cite{NKGS,diss} for
heavy quark production through light quark annihilation and gluon fusion.
At the level of NLL we can diagonalize the matrix equation and 
calculate the eigenvalues and eigenvectors of the anomalous dimension
matrix. The resummed cross section then becomes a sum of exponentials. 

As mentioned above, we have previously applied these results to top 
and bottom quark production
near partonic threshold in hadronic collisions at a center-of-mass 
scattering angle $\theta=90^\circ$ where the soft anomalous
dimension matrix is diagonal \cite{NKJSRV}. 
Here we consider the general case of the total angle-integrated cross section
where we must diagonalize the matrix and decompose the Born cross section
in a particular color tensor basis. The complications are, as we will 
see below, considerable, even for the $q {\overline q}$ channel.

We now present an analysis of the process
$q(p_a)+{\overline q}(p_b) \rightarrow {\overline Q}(p_1) + Q(p_2)$.
We define a variable $s_4 = s + t_1 + u_1$ in terms of the Mandelstam
invariants $s=(p_a+p_b)^2$, $t_1=(p_a-p_1)^2-m^2$,
and $u_1=(p_b-p_1)^2-m^2$, where $m$ is the heavy quark mass.
The value of $s_4$ depends on the four-momentum $k$ of the radiated gluon
and is given near threshold by $s_4=2m^2(1-z)=2mk^0$.
The $q \overline q$ resummed partonic cross section is then given by
\cite{LSN,HERAB,NKJSRV}
\beqa
\sigma^{\rm res}_{q \overline q}(s,m^2)& = & \sum_{i,j=1}^2\int_{-1}^1 
d\cos \theta \, 
\left[-\int^{s-2ms^{1/2}}_{s_{\rm cut}}
ds_4 \, f_{q \overline q, ij}(s_4,\theta) \,
\frac{d{\overline \sigma}_{q \overline q, ij}^{(0)}
(s,s_4,\theta)}{ds_4}\right] \, ,
\nonumber \\ 
\label{one} 
\eeqa
where $ij$ is the component in color space of the anomalous dimension matrix, 
and $d{\overline \sigma}_{q \overline q, ij}^{(0)}(s,s_4,\theta)/ds_4$ are 
components of the differential of the Born cross 
section defined in the same color basis as the anomalous dimension matrix, 
as explained below.

The function $f_{q \overline q, ij}$ is given at NLL by the exponential 
\beq
f_{q \overline q, ij} \equiv\exp[E_{q \overline q, ij}]
=\exp[E_{q \overline q}+E_{q \overline q}(\lambda_i, \lambda_j)] \label{two}
\eeq
where $\lambda_{i,j}$ are the eigenvalues of the soft anomalous
dimension matrix, $\Gamma_S$.
The universal component, $E_{q \overline q}$, is scheme dependent
and is known from Drell-Yan production \cite{oldDY,CT}.
At NLL in the DIS scheme this universal contribution is given by \cite{NKJSRV}
\begin{eqnarray}
 E_{q\overline q} &=& \int_{\omega_0}^1\frac{d\omega'}{\omega'}
\Big\{\int_{\omega'^2 \mu^2/\Lambda^2}^{\omega' \mu^2/\Lambda^2} 
\frac{d\xi}{\xi}\,
 \Big[ \frac{2 C_F}{\pi} \Big( \alpha_s(\xi) 
\nonumber \\ &&  \qquad \qquad 
+ \frac{K}{2\pi} \alpha^2_s(\xi) \Big) \Big] 
 - \frac{3}{2} \frac{C_F}{\pi} \alpha_s
\Big( \frac{\omega' \mu^2}{\Lambda^2}\Big)
\, \Big\} \, , \label{three}
\end{eqnarray}
where $C_F=(N^2-1)/(2N)$, $C_A=N$, $K=C_A(67/18-\pi^2/6)-5n_f/9$ \cite{KoTr},
$\Lambda$ is the QCD scale parameter, and $\omega_0=s_4/(2m^2)$ with 
$N$ the number of colors and $n_f$ the number of flavors. 
The NLL color-dependent process-specific contribution to the exponent 
is \cite{NKJSRV}
\begin{eqnarray}
E_{q\overline q}(\lambda_i, \lambda_j) =
-\int_{\omega_0}^1\frac{d\omega'}{\omega'}
\left\{\lambda_i \left[\alpha_s \left(\frac{\omega'^2 \mu^2}{\Lambda^2}\right),
\theta \right]
+ \lambda_j^* \left[\alpha_s \left(\frac{\omega'^2 \mu^2}{\Lambda^2}\right),
\theta \right] \right\} \, , \label{four}  
\end{eqnarray}
where $i,j= 1,2$.
The $s_4$ dependence appears only in the argument of the 
coupling constant.
Note that we have introduced a cutoff $s_{\rm cut}$ in the $s_4$
integration in Eq.~(\ref{one}) because, in the exponents, $\alpha_s$ 
diverges when $\omega'^2 \mu^2/\Lambda^2 \sim 1$, corresponding to a minimum
$s_4$ of $s_{4, {\rm min}} =  2m^2 \Lambda/\mu$.  We choose
a value of the cutoff consistent with the sum of the first few terms in the 
perturbative expansion \cite{LSN,distr,HERAB,NKJSRV}, in the
range $30 s_{4, {\rm min}} < s_{\rm cut} < 40 s_{4, {\rm min}}$.
Note that this corresponds to a cutoff on the soft gluon energy of the order
of the decay width of the top, giving a natural boundary
of the nonperturbative region (see also the discussion in \cite{bc}). 
The central value in our cutoff range, $s_{\rm cut} = 35 s_{4, {\rm min}}$, 
corresponds to $s_4/(2m^2)=0.04$ for $m=175$ GeV/$c^2$ 
and $\Lambda=0.2$ GeV.

The anomalous dimension matrix is calculated at partonic threshold, $s_4=0$,
in a color-tensor basis consisting of $s$-channel singlet and octet exchange,
\beqa
c_1=c_{\rm singlet}=\delta_{ab} \, \delta_{12}, \quad \quad 
c_2=c_{\rm octet}=-\frac{1}{2N} c_1 +\frac{1}{2} \delta_{a2} \, 
\delta_{b1} \, . \label{five}
\eeqa
In this basis and in an axial gauge, $A^0=0$, the components of $\Gamma_S$ are
\cite{NKGS}
\begin{eqnarray}
\Gamma_{11}&=&-\frac{\alpha_s}{\pi}C_F(L_{\beta}+1+\pi i),
\nonumber \\
\Gamma_{21}&=&\frac{2\alpha_s}{\pi}
\ln\left(\frac{u_1}{t_1}\right), \;
\Gamma_{12}=\frac{C_F}{2C_A}\Gamma_{21},
\nonumber \\
\Gamma_{22}&=&\frac{\alpha_s}{\pi}\left\{C_F
\left[4\ln\left(\frac{u_1}{t_1}\right)-L_{\beta}-1-\pi i\right]\right.
\nonumber \\ && \quad \quad
\left.\mbox{}+\frac{C_A}{2}\left[-\ln\left(\frac{u_1^2 m^2 s}{t_1^4}\right)
+L_{\beta}+\pi i \right]\right\}\, , \label{six}
\end{eqnarray}
where
\begin{equation}
L_{\beta}=\frac{1-2m^2/s}{\beta}\left(\ln\frac{1-\beta}{1+\beta}
+\pi i \right)\, , \label{seven}
\end{equation}
with $\beta=\sqrt{1-4m^2/s}$. 
Note that the matrix
$\Gamma_S$ is diagonal in this singlet-octet basis when $\beta = 0$ and also
when $\theta=90^\circ$ ($u_1=t_1$) for arbitrary $\beta$.
It is this property that simplifies the calculation of the cross section 
at $\theta=90^\circ$, as shown in Ref. \cite{NKJSRV}.

The eigenvalues $\lambda_i$ and eigenvectors $e_i$, with $i=1,2$, of the 
anomalous dimension matrix are given, respectively, by
\begin{equation}
\lambda_{1,2}=\frac{1}{2}\left[\Gamma_{11}+\Gamma_{22}
\pm \left((\Gamma_{11}-\Gamma_{22})^2+4 \Gamma_{12} \Gamma_{21}\right)^{1/2}
\right] \, , 
\end{equation}
and 
\begin{equation}
e_i=\left[\begin{array}{c}
\frac{\Gamma_{12}}{\lambda_i-\Gamma_{11}} \\ 
1
\end{array}\right] 
\end{equation}
for each eigenvalue $\lambda_i$.
Then, if $C=(c_{\rm singlet}\, , \; c_{\rm octet})$
is the original color basis, the diagonal color basis in which we work is
$C' \equiv (c_1'\, , \; c_2')=CR$, where
\beq
R=[e_1 \; e_2]=\left[\begin{array}{cc}
\frac{\Gamma_{12}}{\lambda_1-\Gamma_{11}} & 
\frac{\Gamma_{12}}{\lambda_2-\Gamma_{11}} \\ 
1 & 1
\end{array}\right] \, ,
\eeq
and the diagonalized anomalous dimension matrix is 
\beq
\Gamma_S^{\rm diag}=R^{-1} \Gamma_S R=\left[\begin{array}{cc}
\lambda_1 & 0 \\
0 & \lambda_2 
\end{array}\right] \, .
\eeq
The diagonal color basis is given explicitly by 
\beq
C' \equiv (c_1' \; , \; c_2')=\left(\frac{\Gamma_{12}}{\lambda_1-\Gamma_{11}}
c_{\rm singlet}+c_{\rm octet} \quad , \quad 
\frac{\Gamma_{12}}{\lambda_2-\Gamma_{11}} c_{\rm singlet}+c_{\rm octet}\right),
\eeq
or, inversely,
$C=C' R^{-1}$, giving for 
the octet contribution 
\beq
c_{\rm octet}=\frac{(\lambda_1-\Gamma_{11}) c_1'
-(\lambda_2-\Gamma_{11})c_2'}{(\lambda_1-\lambda_2)} \, . \label{coct}
\eeq
Since the Born cross section is pure octet exchange, it is proportional
to $c_{\rm octet}^2$ so that, after squaring 
Eq.~(\ref{coct}), the Born cross section is defined
in terms of $|c_1'|^2,|c_2'|^2$, and $c_1'c_2'^*$.
Using the relations $c_{\rm octet}^2=(N^2-1)/4$, $c_{\rm singlet}^2=N^2$,
and $c_{\rm singlet} \, c_{\rm octet}=0$,
we find
\beq
|c_{1,2}'|^2=\frac{N^2 \Gamma_{12}^2}{|\lambda_{1,2}-\Gamma_{11}|^2}
+\frac{N^2-1}{4}
\eeq
and
\beq
c_1' \, c_2'^*=\frac{N^2 \Gamma_{12}^2}{(\lambda_1-\Gamma_{11})
(\lambda_2-\Gamma_{11})^*}+\frac{N^2-1}{4} \, .
\eeq

The resummed partonic cross section can then be written in
the diagonal basis as
\beqa 
&& \hspace{-5mm} 
\sigma^{\rm NLL, res}_{q \overline q}(s,m^2) 
= -\sum_{i,j=1}^2\int_{-1}^1 d\cos \theta \, 
\int^{s-2ms^{1/2}}_{s_{\rm cut}} ds_4 \;
\frac{1}{|\lambda_1-\lambda_2|^2} 
\frac{d{\overline \sigma}_{q \overline q}^{(0)}(s,s_4,\theta)}{ds_4}
\nonumber \\ && \hspace {-5mm} \times
\left[\left(\frac{4N^2}{N^2-1}\Gamma_{12}^2 
+|\lambda_1-\Gamma_{11}|^2 \right) e^{E_{q \overline q, 11}} 
+\left(\frac{4N^2}{N^2-1}\Gamma_{12}^2 
+|\lambda_2-\Gamma_{11}|^2 \right) e^{E_{q \overline q, 22}}\right.
\nonumber \\ && 
\left. -\frac{8N^2}{N^2-1}\Gamma_{12}^2 
{\rm Re}\left(e^{E_{q \overline q, 12}}\right)
-2{\rm Re} \left((\lambda_1-\Gamma_{11})(\lambda_2-\Gamma_{11})^*
e^{E_{q \overline q, 12}}\right)\right] \, ,
\label{twelve}
\eeqa
where
\beqa
&&\frac{d{\overline \sigma}_{q \overline q}^{(0)}(s,s_4,\theta)}{ds_4}=
-\pi \alpha_s^2 K_{q {\overline q}} N C_F 
\frac{1}{4s^4}
\frac{s-s_4}{\sqrt{(s-s_4)^2-4sm^2}}
\nonumber \\ && \quad \times 
\left[(3(s-s_4)^2-8sm^2)
(1+\cos^2\theta)+4sm^2(1-\cos^2\theta)\right]  
\label{thirteen}
\eeqa
is the differential of the Born cross section.
The explicit expressions for all the quantities in Eq.~(\ref{twelve})
are long but straightforward to derive.
For example, if we define variables $r$ and $\phi$ by
\beq
(\Gamma_{11}-\Gamma_{22})^2+4\Gamma_{12}\Gamma_{21}=\frac{\alpha_s^2}{\pi^2}
r e^{i \phi} \, ,
\eeq
we can write the eigenvalues as
\begin{equation}
\lambda_{1,2}=\frac{1}{2}\left[\Gamma_{11}+\Gamma_{22}
\pm \frac{\alpha_s}{\pi} r^{1/2} e^{i \phi/2}\right]  \, . 
\end{equation}
Then, we find for the exponentials
\beqa
e^{E_{q {\overline q}, 11 (22)}}&=&\exp\left\{E_{q \overline q}
-\int_{\omega_0}^1 \frac{d\omega'}{\omega'} \left[
{\rm Re}\Gamma_{11}+{\rm Re}\Gamma_{22} \right. \right.
\nonumber \\ && \left. \left. 
\pm \frac{\alpha_s(\omega'^2 \mu^2/\Lambda^2)}{\pi}
r^{1/2}\cos\left(\frac{\phi}{2}\right)\right]\right\} \, ,
\label{expon11}
\eeqa
where the $+(-)$ sign for the last term is for $E_{q {\overline q}, 11}
(E_{q {\overline q}, 22})$, 
and 
\beqa
{\rm Re} \left(e^{E_{q {\overline q}, 12}}\right)&=&
\exp\left\{E_{q \overline q}
-\int_{\omega_0}^1 \frac{d\omega'}{\omega'} \left[
{\rm Re}\Gamma_{11}+{\rm Re}\Gamma_{22} \right] \right\}
\nonumber \\ &&
\times \cos\left[-\int_{\omega_0}^1\frac{d\omega'}{\omega'} 
\frac{\alpha_s(\omega'^2 \mu^2/\Lambda^2)}{\pi} r^{1/2} 
\sin \left(\frac{\phi}{2}\right)\right] \, . \label{twenty}
\eeqa 
Note that ${\rm Im}(e^{E_{q {\overline q}, 12}})$ is the same 
as ${\rm Re}(e^{E_{q {\overline q}, 12}})$  above but with the
cosine replaced by the sine of the terms 
in the last square brackets of Eq.~(\ref{twenty}).

\mysection{Numerical results}

In this section we present some numerical results for the exponents and the
resummed partonic and hadronic top quark production cross sections.

\begin{figure}[h]
\setlength{\epsfxsize=\textwidth}
\setlength{\epsfysize=0.6\textheight}
\centerline{\epsffile{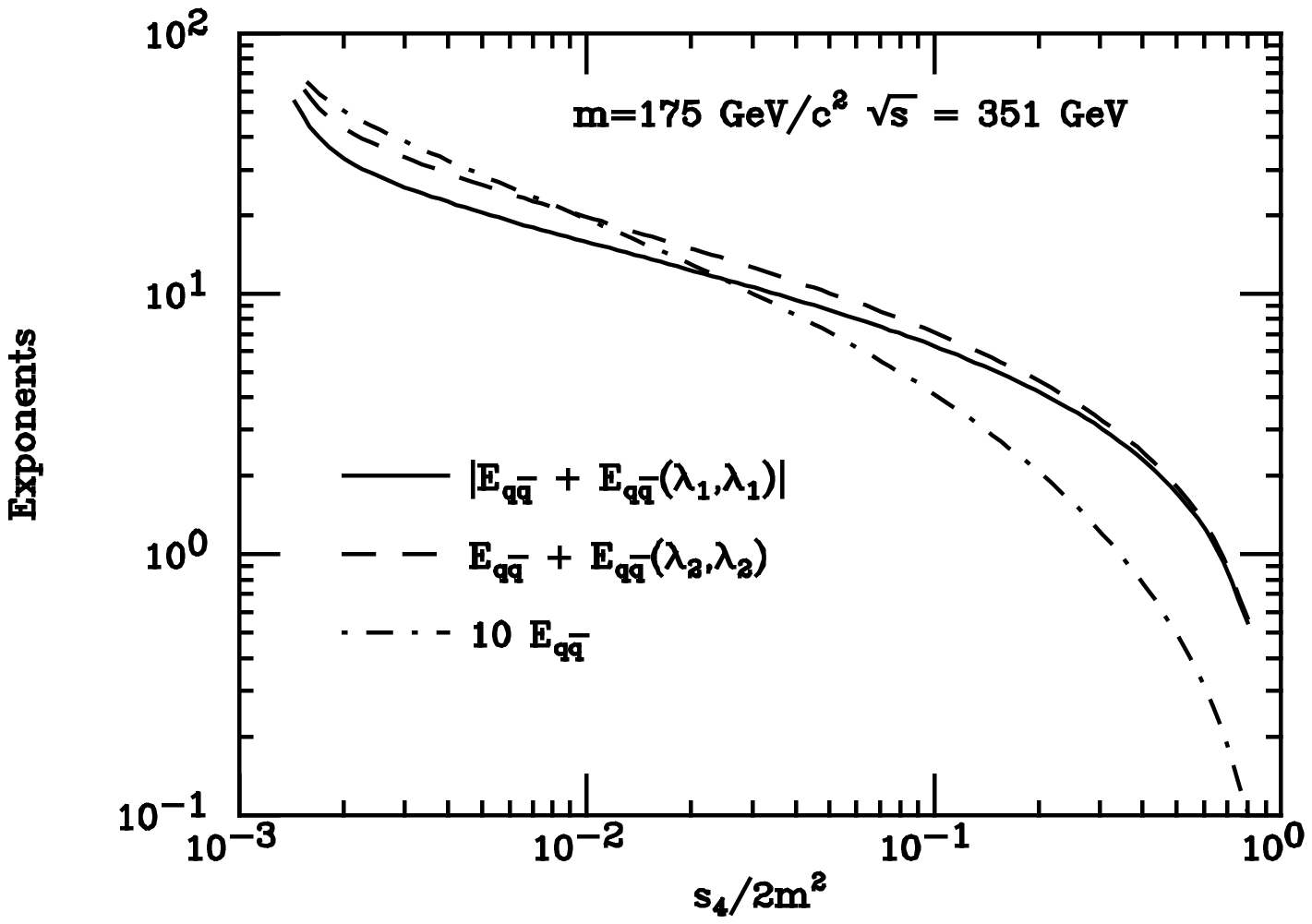}}
\caption {We show the contributions to the $q \overline q$ DIS exponents,
Eqs.~(\ref{two}) and (\ref{expon11}), 
in top quark production for $\mu = m= 175$ GeV/$c^2$ 
and $\sqrt{s} = 351$ GeV as a function of $s_4/(2m^2)$.  The solid curve
shows the absolute value of $E_{q \overline q, 11}$, the dashed curve,
$E_{q \overline q, 22}$. The dot-dashed curve shows $10 E_{q \overline q}$,
Eq.~(\ref{three}). The factor of 10 is included to facilitate comparison.}
\label{figone}
\end{figure}

In Fig.~\ref{figone}, the exponents in the resummed cross section,
$E_{q {\overline q},11}$ and $E_{q {\overline q},22}$, Eqs.~(\ref{two})
and (\ref{expon11}), along
with the universal contribution, $E_{q \overline q}$, Eq.~(\ref{three}), are
given as functions of $s_4/(2m^2)$ in the DIS scheme
with $m=175$ GeV/$c^2$, $\sqrt s=351$ GeV, and $\Lambda_5 = 0.202$ GeV to be
consistent with the CTEQ 4D parton densities \cite{CTEQ,PDF}.  
Note that the absolute value of $E_{q \overline q,11}$ is given.
The color-dependent exponents
are considerably larger than $E_{q \overline q, 22}$ 
at fixed angle, as shown in Fig. 1(a) of the first reference of \cite{NKJSRV}.

\begin{figure}[h]
\setlength{\epsfxsize=\textwidth}
\setlength{\epsfysize=0.6\textheight}
\centerline{\epsffile{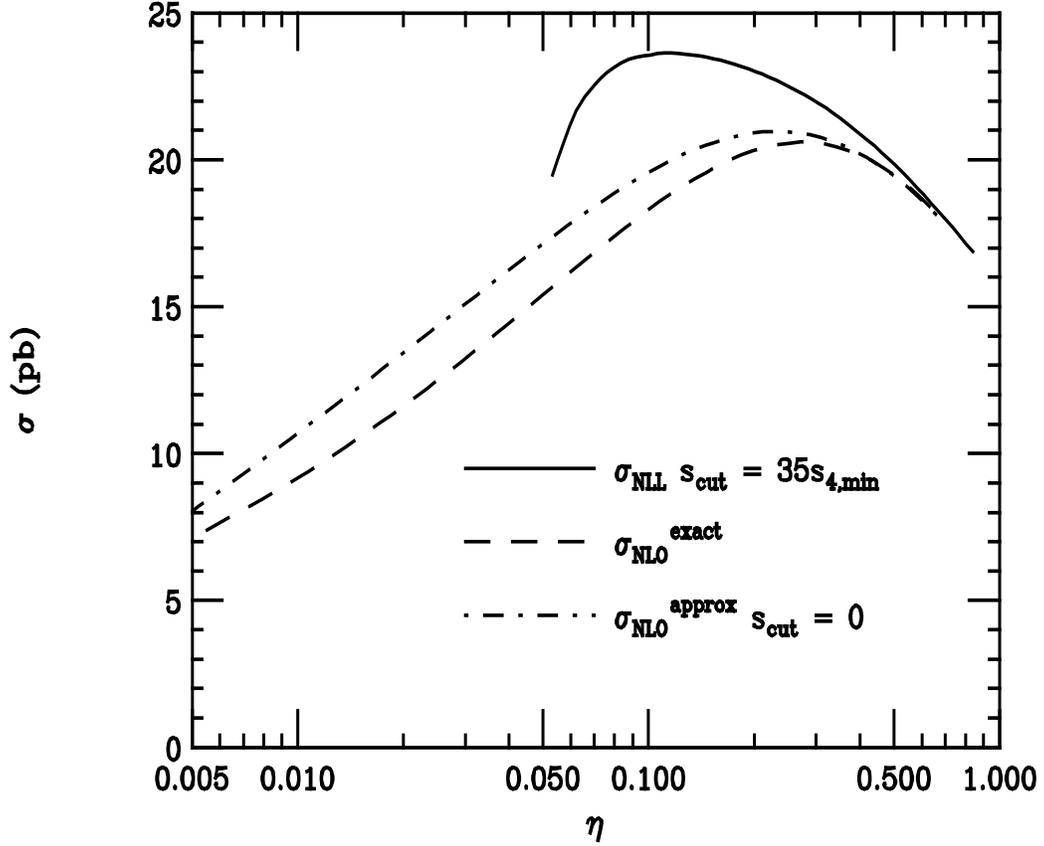}}
\caption {The resummed partonic top quark production cross section,
Eq.~(\ref{twelve}), is shown (solid curve) 
as a function of $\eta= s/(4m^2)-1$ 
for the $q \overline q$ channel in the DIS scheme and $\mu = m = 175$ GeV/$c^2$
with $s_{\rm cut}= 35 s_{4, {\rm min}}$.  
The NLO exact (dashed) and approximate, with $s_{\rm cut}=0$, 
(dot-dashed) cross sections are also given.} 
\label{figtwo}
\end{figure}

In Fig.~\ref{figtwo}, the partonic top quark cross section, 
Eq.~(\ref{twelve}), is presented as a function
of $\eta=s/(4m^2)-1$.  We show the NLO exact and approximate results as well as
the NLL resummed result (with $s_{\rm cut}=35s_{4,{\rm min}}$).
The NLO approximate cross section is the one-loop expansion of our
NLL resummed cross section.
The lower limit of the $\eta$ range of the NLL resummed partonic 
cross section depends on the value of $s_{\rm cut}$.  
No cutoff is applied to the NLO approximate results.
As shown in \cite{NKGS,diss}, the one-loop expansion of the resummed 
cross section agrees analytically near threshold with the NLO approximate 
results in \cite{Mengetal}, and, as is evident from Fig.~\ref{figtwo}, 
is an excellent approximation 
of the exact NLO cross section in the threshold region 
(see also Fig. 7 in \cite{Mengetal}).
We note that the largest contribution to the hadronic cross section
comes from the region $0.1 < \eta < 1$ \cite{HERAB,diss}. 

The NLL resummed hadronic cross section is given by the convolution
of parton distributions $\phi_{i/h}$, for parton $i$ in hadron $h$,
with the partonic cross section 
\begin{eqnarray}
\sigma^{\rm NLL, res}_{q \overline q, {\rm had}}(S,m^2)&=&\sum_{q=u}^b
\int_{\tau_0}^1 d\tau
\int_\tau^1 \frac{dx}{x}
\phi_{q/h_1}(x,\mu^2) \phi_{{\bar q}/h_2}(\frac{\tau}{x},\mu^2)
\sigma_{q \overline q}^{\rm NLL,res}(\tau S, m^2) \, , 
\nonumber \\ &&
\label{fourteen}
\end{eqnarray}
where $\sigma_{q \overline q}^{\rm NLL,res}(\tau S, m^2)$ is defined in 
Eq.~(\ref{twelve}) and $\tau_0 = (m + \sqrt{m^2 + s_{\rm cut}})^2/S$.  

\begin{figure}[h]
\setlength{\epsfxsize=\textwidth}
\setlength{\epsfysize=0.6\textheight}
\centerline{\epsffile{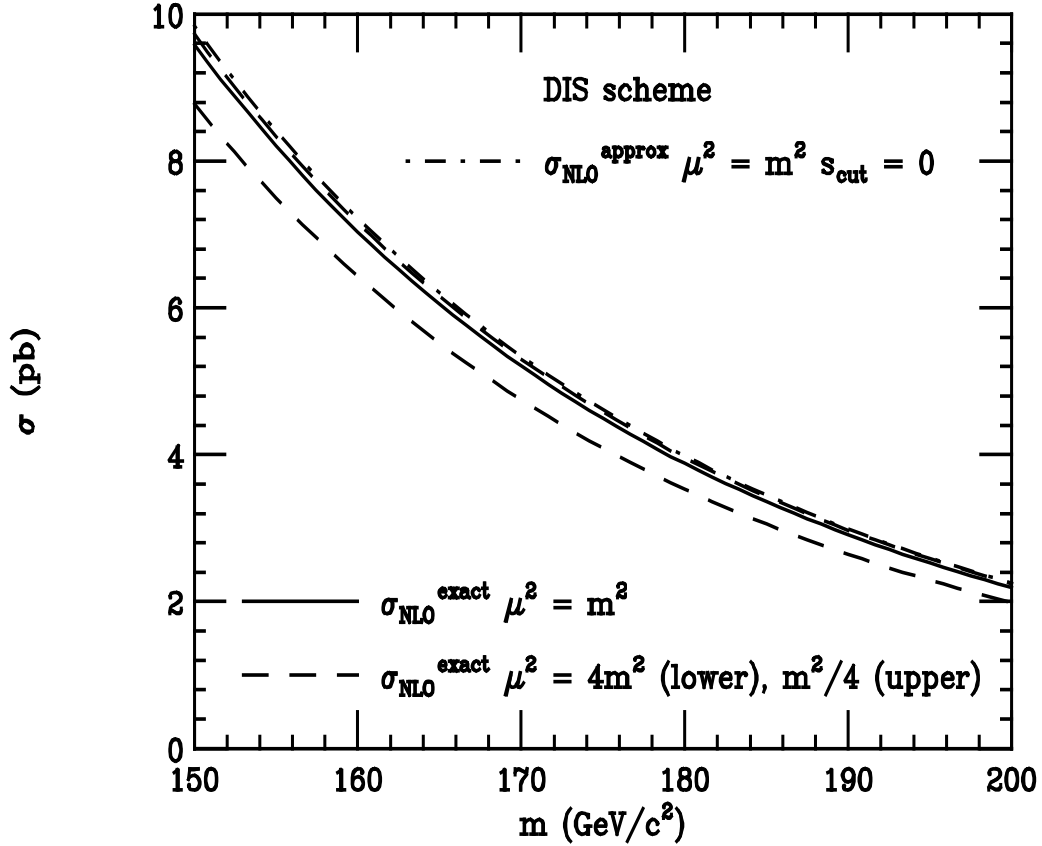}}
\caption {The NLO exact and approximate hadronic $t \overline t$ production 
cross sections in the $q \overline q$ channel and the DIS scheme are
given as functions of top quark mass for $p \overline p$ collisions at the
Tevatron energy, $\sqrt{S} = 1.8$ TeV. The NLO exact cross section is given
for $\mu^2 = m^2$ (solid curve), 
$4m^2$ (lower-dashed) and $m^2/4$ (upper-dashed).  
The NLO approximate cross section with $s_{\rm cut} = 0$ is shown for
$\mu^2 = m^2$ (dot-dashed).}  
\label{figthree_a}
\end{figure}

\begin{figure}[h]
\setlength{\epsfxsize=\textwidth}
\setlength{\epsfysize=0.6\textheight}
\centerline{\epsffile{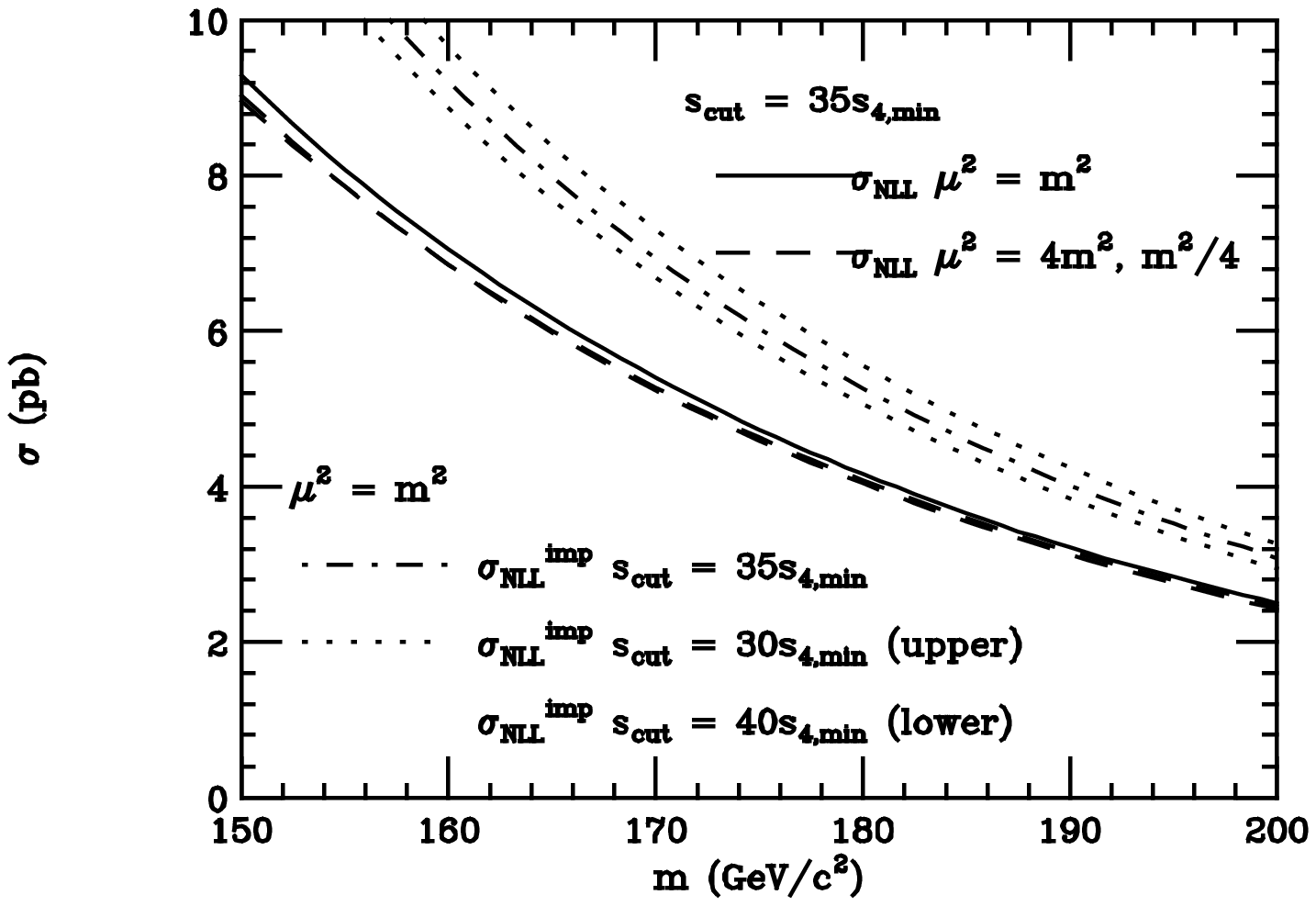}}
\caption {The NLL resummed and improved hadronic $t \overline t$ production 
cross sections in the $q \overline q$ channel and the DIS scheme are
given as functions of top quark mass for $p \overline p$ collisions at the
Tevatron energy, $\sqrt{S} = 1.8$ TeV.  The NLL resummed
cross section, Eq.~(\ref{fourteen}), is shown with $s_{\rm cut} = 
35s_{4,{\rm min}}$ and $\mu^2 = m^2$
(solid curve), $4m^2$ and $m^2/4$ (both dashed).  
The NLL improved cross section,
Eq.~(\ref{improved}), is given for $s_{\rm cut} = 35 s_{4, {\rm  min}}$
(dot-dashed), $30 s_{4, {\rm min}}$ (upper-dotted) and $40s_{4, {\rm min}}$
(lower-dotted).}
\label{figthree_b}
\end{figure}

Our numerical results for the $t \overline t$ production
cross section at the Fermilab Tevatron with $\sqrt S=1.8$ TeV are shown in
Figs.~\ref{figthree_a} and \ref{figthree_b} 
as functions of the top quark mass.
We use the CTEQ 4D DIS parton densities \cite{CTEQ,PDF}.
Since the parton densities are only available at fixed order, 
the application to a resummed cross section
introduces some uncertainty. 
The NLO exact cross sections, including the factorization scale 
dependence, are shown in Fig.~\ref{figthree_a} along with the NLO approximate 
cross section, calculated with $s_{\rm cut}=0$ and $\mu^2=m^2$.  
Again we note the excellent agreement between the NLO exact and
approximate cross sections.

The NLL resummed cross section, including the scale dependence, is shown in
Fig.~\ref{figthree_b}.  The scale dependence is significantly
reduced relative to the NLO cross section. Note that implicit in the
change of scale is a change of cutoff through the definition of 
$s_{4, {\rm min}}$.
To match our results to the exact NLO cross section we define the NLL 
improved cross section 
\beq
\sigma_{q \overline q, {\rm had}}^{\rm imp} 
= \sigma_{q \overline q, {\rm had}}^{\rm NLL, res} -
\sigma_{q \overline q, {\rm had}}^{\rm NLO, approx} 
+ \sigma_{q \overline q, {\rm had}}^{\rm NLO, exact} \, \, . 
\label{improved} 
\eeq
In Fig.~\ref{figthree_b} the hadronic improved cross section 
is shown for $\mu^2 = m^2$ 
along with the variation with $s_{\rm cut}$. 
Note that the improved cross section is calculated with the 
same cut applied to the NLO approximate and the
NLL resummed cross sections.  This cut reduces the approximate cross section
relative to the exact one. The variation of the improved cross section
is small over the range 
$30 s_{4, {\rm min}}< s_{\rm cut} <40 s_{4, {\rm min}}$.

\mysection{Conclusions}

We have given explicit results for the resummed heavy quark
production cross section to next-to-leading logarithmic level.
We have presented numerical results for the dominant channel,
$q \overline q$, in the DIS scheme, for $t \overline t$ production 
at the Tevatron. At $m = 175$ GeV/$c^2$ and $\sqrt{S} = 1.8$ TeV, 
the value of the improved cross section for 
$q \overline q \rightarrow t \overline t$ with 
$s_{\rm cut}/(2m^2)=0.04$ is 
$5.7 < \sigma_{q \overline q, {\rm had}}^{\rm imp} <6.1$ pb
for $m/2 < \mu < 2m$, with a central value of 6.0 pb 
compared to a NLO cross section of 4.5 pb at $\mu=m$.
At the upgraded Tevatron with $\sqrt{S} = 2$ TeV, the corresponding
range is  $7.4 < \sigma_{q \overline q, {\rm had}}^{\rm imp} <7.9$ pb,  
with a central value of 7.8 pb compared to a NLO cross section of
5.9 pb at $\mu=m$. We find that 
the corrections relative to NLO are larger for larger scales. 
The $gg$ channel is more complicated and a complete analysis
will be given elsewhere \cite{NKRV}.  
Adding the $gg$ contribution
we predict a total cross section of 7 pb at $\sqrt{S} = 1.8$ TeV, 
in good agreement with experimental values from CDF, 
$\sigma_{t \overline t}=7.6^{+1.8}_{-1.5}$ pb \cite{CDF}, and D0, 
$\sigma_{t \overline t}=5.5 \pm 1.8$ pb \cite{D0}.  
Gluon fusion is more important
for $b$-quark production at HERA-B where threshold resummation
is also of importance \cite{HERAB}.
Our formalism can also be naturally extended to heavy quark inclusive 
differential distributions in transverse momentum and rapidity \cite{distr}.

\mysection*{Acknowledgements}

The work of N.K. was supported by the PPARC under grant GR/K54601.
The work of R.V. was supported in part by
the Director, Office of Energy Research, Division of Nuclear Physics
of the Office of High Energy and Nuclear Physics of the U. S.
Department of Energy under Contract Number DE-AC03-76SF0098.
We wish to thank Eric Laenen, Sven Moch, Gianluca Oderda, Jack Smith, and 
George Sterman for many helpful conversations.

\pagebreak

\pagebreak

\pagebreak

\pagebreak

\begin{thebibliography}{99}

\bibitem{CSS} J.C.\ Collins, D.E.\ Soper, and G.\ Sterman,
in {\it Perturbative Quantum Chromodynamics},
ed.\ A. H.\ Mueller (World Scientific, Singapore, 1989), p. 1.

\bibitem{heavycalcs}  P.\ Nason, S.\ Dawson, and R.K.\ Ellis,
Nucl.\ Phys.\ {\bf B303}, 607 (1988); 
W.\ Beenakker, H.\ Kuijf, W.L.\ van Neerven,
and J.\ Smith Phys.\ Rev.\ D {\bf 40}, 54 (1989); 
W.\ Beenakker, W.L.\ van
Neerven, R.\ Meng, G.A.\ Schuler, and J.\ Smith, Nucl.\ Phys.\ 
{\bf B351}, 507 (1991).

\bibitem{Thuile} N. Kidonakis and G. Sterman, 
in {\sl Les Rencontres de Physique de
la Vall\'ee d'Aoste, Results and Perspectives in Particle Physics}, 
ed.\ M.\ Greco (INFN-Frascati SIS-Ufficio 
Pubblicazioni, Italy, 1996), p.\ 333, hep-ph/9607222.

\bibitem{CLS} H. Contopanagos, E. Laenen, and G. Sterman,
Nucl.\ Phys.\ {\bf B484}, 303 (1997).

\bibitem{oldDY} G.\ Sterman, Nucl.\ Phys.\ {\bf B281}, 310 (1987).

\bibitem{CT}
 S.\ Catani and L.\ Trentadue,
Nucl.\ Phys.\ {\bf B327}, 323 (1989);
{\bf B353}, 183 (1991).

\bibitem{LSN} E. Laenen, J. Smith, and W.L. van Neerven,
Nucl. Phys. {\bf B369}, 543 (1992); Phys. Lett. B {\bf 321}, 254 (1994).

\bibitem{distr} N. Kidonakis and J. Smith, Phys. Rev. D {\bf 51}, 6092 (1995).

\bibitem{HERAB}
N. Kidonakis and J. Smith, Mod. Phys. Lett. A {\bf 11}, 587 (1996);
hep-ph/9506253;
J. Smith and R. Vogt, Z. Phys. C {\bf 75}, 271 (1997).

\bibitem{bc} E.L.\ Berger and H.\ Contopanagos, Phys. Lett. B {\bf 361},
115 (1995); Phys. Rev. D {\bf 54}, 3085 (1996); {\it ibid.} D {\bf 57}, 
253 (1998).

\bibitem{cat} S. Catani, M.L. Mangano, P. Nason, and L. Trentadue,
Nucl. Phys. {\bf B478}, 273 (1996); Phys. Lett. B {\bf 378}, 329 (1996).

\bibitem{NKGS} N. Kidonakis and G. Sterman, Phys. Lett. B {\bf 387}, 867
(1996); Nucl. Phys. {\bf B505}, 321 (1997); in proceedings of {\it DIS 97},
ed. J. Repond and D. Krakauer (AIP Conf. Proc. No. 407, Woodbury, NY, 1997),
p. 1035, hep-ph/9708353.

\bibitem{diss} N. Kidonakis, Ph.D. Thesis, 1996, hep-ph/9606474.

\bibitem{KOS} N.\ Kidonakis, G.\ Oderda, and G.\ Sterman, 
Nucl. Phys. {\bf B525}, 299 (1998); hep-ph/9803241,
to appear in Nucl. Phys. {\bf B}; 
talk presented at {\it DIS 98}, Brussels, April 4-8, 1998, hep-ph/9805279.

\bibitem{LOS} E. Laenen, G.\ Oderda, and G.\ Sterman, hep-ph/9806467.

\bibitem{MoLa} E. Laenen and S. Moch, hep-ph/9809550;
S. Moch, talk presented at {\it DIS 98}, 
Brussels, April 4-8, 1998, hep-ph/9805370.

\bibitem{OS} G. Oderda and G. Sterman, Phys. Rev. Lett. 81, 3591 (1998);
G. Oderda, talk presented at {\it QCD 98}, Montpellier, July 2-8, 1998,
hep-ph/9808384.

\bibitem{BonCat} R. Bonciani, S. Catani, M.L. Mangano, and P. Nason,
Nucl. Phys. {\bf B529}, 424 (1998).

\bibitem{CMN} S. Catani, M.L. Mangano, and P. Nason,
J. High Energy Phys. 9807, 024 (1998).

\bibitem{NKJSRV} N. Kidonakis, J. Smith, and R. Vogt, Phys. Rev. D {\bf 56},
1553 (1997); N. Kidonakis, in {\it QCD 97}, Nucl. Phys.  
(Proc. Suppl.) {\bf B64}, 402 (1998).

\bibitem{NKRV} N. Kidonakis and R. Vogt, in preparation.

\bibitem{KoTr} J. Kodaira and L. Trentadue, Phys. Lett. B {\bf 112}, 66 (1982).

\bibitem{CTEQ} H.L. Lai, J. Huston, S. Kuhlmann, F. Olness, J. Owens, 
D. Soper, W.K. Tung, and H. Weerts, Phys. Rev. D {\bf 55}, 1280 (1997).

\bibitem{PDF} H. Plothow-Besch,  `PDFLIB: Nucleon, Pion
and Photon Parton Density Functions and $\alpha_s$ Calculations',
Users's Manual - Version 7.09, W5051 PDFLIB, 1997.07.02, CERN-PPE.

\bibitem{Mengetal} R.\ Meng, G.A.\ Schuler, J.\ Smith, and W.L.\ van
Neerven, Nucl.\ Phys.\ {\bf B339}, 325 (1990).

\bibitem{CDF} F. Abe {\it et al.} (CDF Collab.),
Phys. Rev. Lett. {\bf 80}, 2773 (1998).

\bibitem{D0} S. Abachi {\it et al.} (D0 Collab.),
Phys. Rev. Lett. {\bf 79}, 1203 (1997). 
 
\end{thebibliography}
\end{document}